\documentclass[a4paper]{JHEP3}
\usepackage[latin1]{inputenc}
\usepackage{amssymb}
\usepackage{slashed}
\usepackage[english]{babel}
\usepackage{feynmp}
\usepackage{epsfig}
\usepackage{cite}
\usepackage{mcite}

\newcommand{\half}{\frac{1}{2}}
\newcommand{\Tr}{\mathrm{Tr}}
\newcommand{\mE}{m_\mathrm{E}}
\newcommand{\gE}{g_\mathrm{E}}
\newcommand{\MSbar}{\overline{\textrm{MS}}}
\newcommand{\dd}{\mathrm{d}}
\newcommand{\xbot}{\mathbf{x}_\bot}
\newcommand{\Ei}{\mathrm{Ei}}
\newcommand{\CF}{C_\mathrm{F}}
\def\sumint{\hbox{$\sum$}\!\!\!\!\!\!\!\int}
\newcommand{\pnij}{\bar{p}_{0\mathrm{ij}}}
\newcommand{\Mij}{\bar{M}_\mathrm{ij}}

\title{Mesonic screening masses at high temperature and finite density}

\author{Mikko Veps\"al\"ainen\\
Theoretical Physics Division,
Department of Physical Sciences,\\
P.O.Box 64, FIN-00014 University of Helsinki, Finland\\
E-mail: \email{Mikko.T.Vepsalainen@helsinki.fi} }

\abstract{
We compute the first perturbative correction to the static correlation 
lengths of light quark bilinears in hot QCD with finite quark chemical 
potentials. The correction is small and positive, with $\mu$-dependence 
depending on the relative sign of chemical potentials and the number of 
dynamical flavors. The computation is carried out using a 
three-dimensional effective theory for the lowest fermionic Matsubara 
mode. We also compute the full correlator in free theory and find a 
rather complicated general $\mu$-dependence at shorter distances. 
Finally, rough comparisons with lattice simulations are discussed.}

\preprint{hep-ph/0701250}

\begin{document}

\begin{fmffile}{diagrams}

\section{Introduction}

With the ongoing and future heavy ion experiments at RHIC and LHC, we have at hand a large body of experimental data on QCD at extreme conditions. However, comparing this data with the theory predictions is far from trivial, because we cannot usually reliably compute the measured physical quantities directly from QCD, even in the case when the plasma reaches an equilibrium.

The only systematic non-perturbative approach is lattice gauge field theory in four dimensions. With an increasing amount of computer power available, it has become possible to determine many quantities starting from first principles, while avoiding the problems arising from the strong coupling and higher order infrared divergences that make perturbation theory so hard near the phase transition. Lattice simulations, however, have potentially large systematic errors arising from light dynamical quark degrees of freedom. In the presence of chemical potential, which is the situation studied in this paper, the situation is even worse, since the quark determinant becomes complex and conventional importance sampling does not work.

There is clearly room for other methods, whose results, even if they have their own limitations, could be compared with the lattice to gain insight on the shortcomings of both. Perturbation theory suffers from infrared problems at high orders \cite{Linde:1980ts,*Gross:1981br}, but these are only related to static, soft gluons. This has led to a technique called dimensional reduction, where the nonperturbative infrared behavior is pushed to a simpler effective theory for soft modes only, allowing the lower order contributions to be computed reliably in perturbation theory. Of course, at moderate temperatures the QCD coupling is large and higher order terms cannot be neglected, but in this way computationally intensive nonperturbative methods can be applied to a simpler 3d theory, while the terms with dynamical quarks and quark chemical potentials can be computed perturbatively in four dimensions. On the other hand, at very high temperatures the perturbation theory result computed in this way should be reliable because of the asymptotic freedom.

In the bosonic sector dimensional reduction has been used successfully in both QCD and electroweak theory to compute quantities like pressure \cite{Kajantie:1995dw,Braaten:1996jr,Kajantie:2002wa,*Kajantie:2003ax,Blaizot:2003iq, Vuorinen:2003fs,Ipp:2006ij,Gynther:2005dj,*Gynther:2005av} and various gluonic correlators \cite{Reisz:1992er,*Karkkainen:1992jh,*Karkkainen:1993wu,Kajantie:1997tt,Laine:1998nq,*Laine:1999hh,Hart:1999dj,Hart:2000ha,Cucchieri:2001tw}. Fermionic modes, on the other hand, are usually integrated out from the effective theory and only affect the parameters of the dimensionally reduced theory. However, there are many interesting observables constructed of quark fields which are sensitive to infrared physics and whose perturbative treatment thus requires some kind of resummations, conveniently organized as successive effective theories. In this paper we concentrate on the long distance behavior of correlators of mesonic operators composed of light (massless) quark flavors \cite{DeTar:1987ar,*DeTar:1987xb}.

The first efforts on treating this kind of correlators in a dimensionally reduced theory date back almost 15 years \cite{Hansson:1992kb,*Hansson:1994nb,Koch:1992nx,*Koch:1994zt}. However, those works did not systematically include all terms of a given order. In \cite{Huang:1996tz} it was shown that the dimensionally reduced theory can be formulated in terms of nonrelativistic quarks, and the correct power counting of different operators was established. With similar methods we  \cite{Laine:2003bd} were able to consistently compute the $\mathcal{O}(g^2)$ corrections to the screening masses of mesonic operators. The current paper aims to extend that work for situations with nonzero quark chemical potentials. These masses have been measured recently on lattice both at zero \cite{Wissel:2005pb,Gavai:2006fs} and finite density \cite{Pushkina:2004wa}. Near the QCD phase transition these correlators have also been studied at finite temperature and density using the Nambu--Jona-Lasinio model \cite{Hansen:2006ee}.

This paper is organized as follows. In Sec.~\ref{sec:correlators} we define the correlators that we intend to study and compute the leading order (free theory) results. The derivation of the effective theory along the lines of \cite{Laine:2003bd} is carried out in Sec.~\ref{sec:et}, as well as the matching of the parameters to the level required for computing the screening masses. In Sec.~\ref{sec:solution} we formulate the problem in the context of the effective theory and solve the screening mass numerically.
Discussion of the results and comparison with lattice data can be found in their respective subsections. Our conclusions are presented in Sec.~\ref{sec:conclusions}. Appendices \ref{app:freetheory} and \ref{app:potential} contain the details of computing the leading order correlators and quark-antiquark potential, respectively.

\section{Static correlators in high temperature QCD}
\label{sec:correlators}

The quark sector in finite temperature QCD is described by the Euclidean Lagrangian
\begin{equation}
 \mathcal{L}_\mathrm{E}^\mathrm{q} = \bar{\psi}(\gamma_\mu D_\mu + M)\psi,
\end{equation}
where the covariant derivative of quark field is defined as
\begin{equation}
 D_\mu \psi \equiv \partial_\mu \psi -ig A_\mu^a T^a \psi,
\end{equation}
with $T^a$ the generators of the fundamental representation of SU($N_c$). The quark field $\psi$ is an $N_F$-component vector
in flavor space, and the $N_F \times N_F$ mass matrix $M$ is taken to vanish in most of what follows.

Finite quark densities can be introduced via chemical potentials $\mu_\mathrm{f}$. Inserting $\mu_\mathrm{f}$ into the path integral leads
to the effective Lagrangian (with $M=0$)
\begin{equation}
 \mathcal{L}_\mathrm{q} = \sum_\mathrm{f} \bar{\psi}_\mathrm{f} (\gamma_\mu D_\mu + \mu_\mathrm{f} \gamma_0) \psi_\mathrm{f}.
 \label{eq:L_QCD}
\end{equation}
In perturbation theory the effect of the chemical potential is then to give a constant imaginary part to the momentum zero-component, $p_0 \to p_0 - i\mu_\mathrm{f}$. The chemical potentials for each quark flavor can be chosen independently when no weak interactions are present, while weak interactions would only conserve certain specific combinations of baryon and lepton numbers, see e.g.~\cite{Gynther:2003za}. We try to give our results for general $\mu_\mathrm{f}$ when possible, but will mainly use isoscalar ($\mu_\mathrm{u}=\mu_\mathrm{d}\equiv \mu_S$) and isovector ($\mu_\mathrm{u}=-\mu_\mathrm{d} \equiv \mu_V$) chemical potentials to discuss the effects of various combinations of $\mu_\mathrm{f}$.

With the quark fields we can define various bilinear objects of different spin and flavor structure,
\begin{equation}
 O^a = \bar{\psi} F^a \Gamma \psi,
\label{eq:general_bilinear}
\end{equation}
where $\Gamma$ is one of $\{ 1,\gamma_5,\gamma_\mu,\gamma_\mu \gamma_5 \}$ for scalar, pseudoscalar, vector and axial vector objects $S^a$, $P^a$, $V_\mu^a$ and $A_\mu^a$, respectively. The identity matrix $F^s$ and the traceless matrices $F^a$ generate the flavor basis,
\begin{equation}
 F^s \equiv 1_{N_F}, \quad \Tr[F^a F^b] = \half \delta^{ab}, \quad a,b=1,\ldots,N_F^2-1.
\end{equation}

We are interested in correlators of the form
\begin{equation}
 C_\mathbf{q}[ O^a, O^b] \equiv \int_0^{1/T}\! \dd \tau \int \! \dd^3 x \, e^{i\mathbf{q\cdot x}}
\langle O^a(\tau,\mathbf{x}) O^b(0,\mathbf{0})\rangle,
\end{equation}
or, in configuration space,
\begin{equation}
 C_\mathbf{x}[ O^a, O^b ] \equiv \int_0^{1/T}\! \dd \tau
\langle O^a(\tau,\mathbf{x}) O^b(0,\mathbf{0})\rangle.
\end{equation}

Because of rotational invariance, the general structure of correlators in terms of the three-momentum $\mathbf{q}$ and the chemical potentials $\mu_\mathrm{f},\mu_\mathrm{g}$ of the quark flavors in the operator~is
\begin{eqnarray}
 \lefteqn{ C_\mathbf{q}[S^a, S^b],\  C_\mathbf{q}[V^a_0, V^b_0] = \sum_\mathrm{fg} F^a_\mathrm{fg}F^b_\mathrm{gf}\, f(q^2,\mu_\mathrm{f},\mu_\mathrm{g}), } \\
 C_\mathbf{q}[V^a_i, V^b_j] &=& \sum_\mathrm{fg} F^a_\mathrm{fg}F^b_\mathrm{gf}\left[ \left( \delta_{ij} 
	-\frac{q_i q_j}{q^2}\right) t(q^2,\mu_\mathrm{f},\mu_\mathrm{g}) 
	+\frac{q_i q_j}{q^2}l(q^2,\mu_\mathrm{f},\mu_\mathrm{g}) \right],
\end{eqnarray}
and similarly for $P^a$ and $A_\mu^a$. Here $q \equiv |\mathbf{q}|$. We expect that the functions $f$, $t$ and $l$ are dominated by simple poles near the origin $q^2=0$, corresponding to a set of bound states in a $2+1$-dimensional theory obtained by analytic continuation. Their Fourier transforms then exhibit exponential falloff at large distances, and the coefficients of this falloff are referred to as screening masses. 

Without a loss of generality we can select the direction of $\mathbf{x}$ as the $x_3$-direction. We make this choice and further average over the $x_1 x_2$-plane, giving a more easily calculable correlator
\begin{equation}
 \label{eq:pointplanedef}
 C_z[O^a, O^b] = \int \! \dd^2 \xbot C_{(\xbot,z)} [O^a, O^b] = \int_0^{1/T}\!\dd \tau \int \! \dd^2 \xbot 
	\langle O^a(\tau,\xbot,z) O^b(0,\mathbf{0},0) \rangle.
\end{equation}

\subsection{Leading order in perturbation theory}

At very high temperatures the strong coupling constant approaches zero because of asymptotic freedom. Perturbation theory should then be applicable, and we can expand the result in powers of $\alpha_s$ around the perturbative vacuum. The leading order result for the correlator in Eq.~(\ref{eq:pointplanedef}) is given by the free theory one loop diagram of Fig.~\ref{fig:diagram_types}(a).

\FIGURE[t]{
\parbox{0.9\textwidth}{
\begin{center}
\begin{fmfgraph*}(60,30)
 \fmfleft{l}
 \fmfright{r}
 \fmf{plain,right=0.5,label=(a)}{l,r}
 \fmf{plain,right=0.5}{r,l}
 \fmfv{d.sh=square,d.size=5,d.fill=empty}{l,r}
\end{fmfgraph*}
\hspace{1cm}
\begin{fmfgraph*}(60,30)
 \fmfleft{l}
 \fmfright{r}
 \fmf{plain,right=0.5,label=(b)}{l,r}
 \fmf{plain,right=0.5}{r,l}
 \fmfv{d.sh=square,d.size=5,d.fill=empty}{l,r}
 \fmffreeze
 \fmfipath{p[]}
 \fmfiset{p1}{vpath(__r,__l)}
 \fmfiset{p2}{vpath(__l,__r)}
 \fmfi{photon}{point length(p1)/2 of p1 -- point length(p2)/2 of p2}
\end{fmfgraph*}
\hspace{1cm}
\begin{fmfgraph*}(60,30)
 \fmfleft{l}
 \fmfright{r}
 \fmf{plain,right=0.5,label=(c)}{l,r}
 \fmf{plain,right=0.5}{r,l}
 \fmfv{d.sh=square,d.size=5,d.fill=empty}{l,r}
 \fmffreeze
 \fmfipath{p}
 \fmfipair{v[]}
 \fmfiset{p}{vpath(__r,__l)}
 \fmfiset{v1}{point length(p)/4 of p}
 \fmfiset{v2}{point 3length(p)/4 of p} 
 \fmfi{photon}{v1{dir -135}..v2{dir135} }
\end{fmfgraph*}
\end{center}}
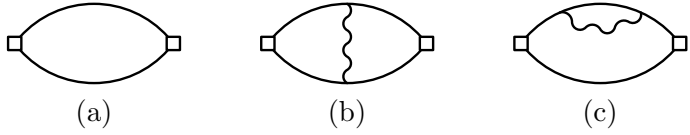
\caption{Diagrams contributing to the meson correlator.}
\label{fig:diagram_types} }

For isoscalar chemical potential the leading order correlator can be computed exactly. A straightforward calculation, the details of which can be found in Appendix \ref{app:freetheory}, gives
\begin{eqnarray}
\label{eq:vapaaskalaari}
 C_z[S^a, S^b] &=& \delta^{ab} \frac{N_c T}{8\pi z \sinh 2\pi T z}\left( 2\pi T \coth 2\pi Tz \, \cos 2\mu z 
	+2\mu \sin 2\mu z +\frac{1}{z} \cos 2\mu z \right) \qquad \\
 &=& \delta^{ab} \frac{N_c}{8\pi^2}\frac{1}{z^3}\left[ 1 -\left( \frac{7}{360} +\frac{1}{6}\frac{\mu^2}{\pi^2 T^2}
	+\frac{1}{12}\frac{\mu^4}{\pi^4 T^4}\right)(2\pi Tz)^4 +\mathcal{O}(z^5) \right] \nonumber \\
 &=& \delta^{ab} \frac{N_c T^2}{2}\frac{1}{z}e^{-2\pi Tz} \left[ \left(1+\frac{1}{2\pi Tz}\right)\cos 2\mu z 
	+\frac{\mu}{\pi T} \sin 2\mu z \right] +\mathcal{O}(e^{-4\pi Tz}), \nonumber
\end{eqnarray}
which for $\mu=0$ agrees with the previous zero density computation by Florkowski and Friman \cite{Florkowski:1993bq}.
In Fig.~\ref{fig:vapaa} we have plotted the expression in Eq.~(\ref{eq:vapaaskalaari}) for various $\mu$. The wavelength of the oscillation from terms like $\cos(2\mu z)$ is $l_\mu =\pi/\mu$, while the exponential falloff is characterized by the screening length $\xi^{-1} = 2\pi T$. When $l_\mu \gg \xi$, or $\mu \ll 2\pi^2 T$, the oscillations become important at distances where the correlator is exponentially suppressed by screening and show only as small ripples in the tail part of the $C_z$. For larger chemical potentials there are strong oscillations inside the exponentially decaying envelope.

This behavior is verified by Fig.~\ref{fig:vapaa}, where the oscillatory behavior cannot really be seen for $\mu \lesssim 0.5 \pi T$, but the correlator just falls off faster than for zero density. At short distance this can be viewed as increased screening, but we will define the screening mass strictly as the coefficient of the exponential falloff, or the real part of the pole location. Nevertheless, the imaginary parts will be computed as well.

\EPSFIGURE[t]{vapaa_scaled.eps,width=\textwidth}{
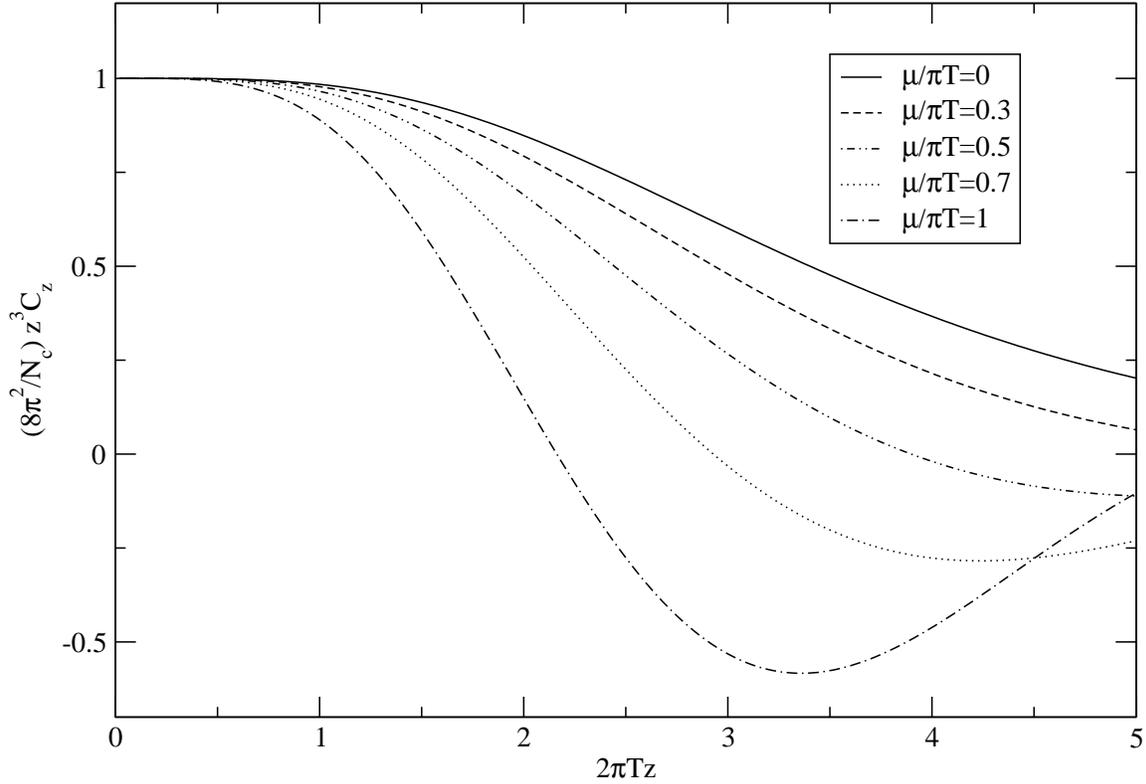
The free correlator of Eq.~(\ref{eq:vapaaskalaari}) for various values of the isoscalar chemical potential.
\label{fig:vapaa}}

The free correlator with arbitrary chemical potentials is much harder to compute. In Appendix \ref{app:freetheory} we compute an approximate correlator by restricting the sum over Matsubara modes to $\omega_n^f = \pm \pi T$, since they dominate the long distance behavior of the correlator. The result can be given in terms of the exponential integral function $\Ei(x)$, and to leading order in $1/z$ it is
\begin{eqnarray}
 C_z[S^a, S^b] &=& \sum_\mathrm{ij} F^a_\mathrm{ij}F^b_\mathrm{ji} \frac{N_c T^2}{2\pi}\left( 1 -e^{-\Delta\mu_\mathrm{ij}/T} \right)
	\frac{1}{z}e^{-2\pi Tz}\left[ \left(\frac{1}{\Delta\mu_\mathrm{ij}} - \frac{\Delta\mu_\mathrm{ij}}{(2\pi T)^2 +
	\bar{\mu}_\mathrm{ij}^2}\right)\bar{\mu}_\mathrm{ij} \sin \bar{\mu}_\mathrm{ij}z \right. \nonumber \\
 &&{}\left. +\left(\frac{1}{\Delta\mu_\mathrm{ij}} + \frac{\Delta\mu_\mathrm{ij}}{(2\pi T)^2
	+\bar{\mu}_\mathrm{ij}^2}\right)2\pi T \cos \bar{\mu}_\mathrm{ij}z \right],
\label{eq:vapaayleinen_appr}
\end{eqnarray}
where $\Delta\mu_\mathrm{ij}\equiv \mu_\mathrm{i}-\mu_\mathrm{j}$ and $\bar{\mu}_\mathrm{ij}\equiv \mu_\mathrm{i}+\mu_\mathrm{j}$. It should be noted that at this order the scale of the oscillation is $\bar{\mu}_\mathrm{ij}$, which vanishes for isovector chemical potential. The subleading terms contain oscillations with $\Delta\mu_\mathrm{ij}$, as can be seen in Eq.~(\ref{eq:vapaayleinen}).

Pseudoscalar correlators differ only by a sign from Eqs.~(\ref{eq:vapaaskalaari}) and (\ref{eq:vapaayleinen_appr}). The (axial) vector correlators can be likewise computed in the free theory and have the same general behavior $C_z \sim \cos(\bar{\mu}_\mathrm{ij}z) \exp(-2\pi T z)$, while the structure is a bit more complicated.


\section{Effective theory}
\label{sec:et}

We would like to compute the next-to-leading order correction to the screening mass in perturbation theory. As usual in finite temperature field theory, simply evaluating the next-to-leading order diagrams in Fig.~\ref{fig:diagram_types}(b,c) is not enough, but we have instead an infinite set of diagrams all contributing at the same order.

First of all, the temporal gluonic zero Matsubara mode $A_0^{(0)}$ gets a self-energy correction of order $g^2T^2$, which for small momenta should be included in the propagator as an electric mass. The introduction of thermal mass resums the diagrams with an arbitrary number of self-energy insertions on the gluon line, and it is required in order to get rid of the infrared divergences that the low momentum electrical gluon modes give rise to in na\"\i ve perturbation theory.

Since we are computing corrections to a mesonic operator near the threshold of producing two free quarks, the quarks can be almost on-shell, with $|1/\slashed{p}| \sim \mathcal{O}(1/g^2 T)$, which compensates for the factors of coupling from the vertices. The diagrams with an arbitrary number of low momentum gluonic zero modes exchanged between quarks should therefore be resummed to get a consistent first order correction to the propagation of free quarks. On the other hand, we do not need to compute the gluon exchange diagrams like in Fig.~\ref{fig:diagram_types}(b) for non-zero gluon modes, since only the one-gluon diagram would contribute at order $g^2$, and this diagram alone cannot change the pole location, \emph{i.e.}\ the screening mass. Of course, if we were computing the overall normalization or the functional form of the correlator at short distances, then these diagrams would contribute.

A convenient way of organizing these resummations is the effective field theory method. The goal is to write a simpler theory for those modes that require special treatment, namely the gluon zero modes and the lowest fermionic modes. These modes reproduce the correct infrared behavior of gauge theory at high temperature \cite{Ginsparg:1980ef,*Appelquist:1981vg}, the relevant scales being of the order $gT$ and $g^2T$, while the contribution of all the other modes can be included in the parameters of the theory. With the sum over Matsubara modes suppressed, the remaining theory is three-dimensional.

In the bosonic sector the dimensional reduction to the physics of gluonic zero modes is well-known. At finite density the Lagrangian has the form \cite{Hart:2000ha}
\begin{equation}
 \mathcal{L}_\mathrm{eff}^\mathrm{b} = \half \Tr F_{ij}^2 + \Tr [D_i,A_0]^2 + \mE^2 \Tr A_0^2 
	+\frac{ig^3}{3\pi^2}\sum_\mathrm{f} \mu_\mathrm{f} \Tr A_0^3
	+ \lambda^{(1)}_\mathrm{E}\left( \Tr A_0^2\right)^2 +\lambda^{(2)}_\mathrm{E}\Tr A_0^4,
\label{eq:bosonicL}
\end{equation}
with $i=1,2,3$, $F_{ij}=i\gE^{-1} [D_i,D_j]$, $D_i=\partial_i-i\gE A_i$ and the parameters to the order we need them are
\begin{equation}
 \mE^2 = g^2 T^2\left( \frac{N_c}{3} +\frac{N_F}{6} + \frac{1}{2\pi^2}\sum_\mathrm{f} \frac{\mu_\mathrm{f}^2}{T^2} \right),
 \qquad \gE^2 = g^2 T,  \qquad \lambda^{(1,2)} = \mathcal{O}(g^4 T). 
\end{equation}
The cubic and quartic $A_0$ interactions can be ignored, since they contribute to meson correlators only at order $g^6$ and higher. The theory in Eq.~(\ref{eq:bosonicL}) describes a 3-dimensional gauge theory with a massive adjoint scalar $A_0^a$.

The fermionic sector consists of the lowest fermionic modes $\omega=\pm \pi T$. The reason we are leaving out the other modes is not that they would be substantially heavier ($\pi T \ll 3\pi T$ is not really true), but because the lowest modes dominate the correlator at large distances. We are expanding around a state consisting of a free quark-antiquark pair, and the relevant expansion parameter is the ``off-shellness'' $p_0^2 +p_\bot^2 +p_3^2$, which near the screening pole $p_3 \sim \pm i(\pi T -i\mu)$ is of the order of the dynamical scales $gT,g^2 T$.

In the three-dimensional theory the lowest fermionic modes can be viewed as massive particles of mass $|\omega_{\pm 1}|=\pi T$. As noted above, the typical momenta are much smaller than this. The correlators then appear as bound states of massive quarks in (2+1)-dimensional theory, a situation resemblant of quarkonia in four dimensions. At $T=0$ the properties of quarkonia can be studied with the help of an effective theory known as non-relativistic QCD (NRQCD) \cite{Caswell:1986ui}, and the fermionic sector of our final effective 3d theory Eq.~(\ref{eq:final_nrqcd}) looks very similar to this theory, although the physical interpretation of the parameters is quite different. Because of similar appearance, we label our effective theory $\textrm{NRQCD}_3$. Its derivation will be presented below.

\subsection{Tree-level $\textrm{NRQCD}_3$}

The dimensional reduction in the fermionic sector using NRQCD methods was first studied by Huang and Lissia in \cite{Huang:1996tz}, where they also computed a number of one-loop corrections to the parameters. A similar derivation in a slightly different formalism was presented in our previous work \cite{Laine:2003bd}. We will here review the reduction in the case of finite chemical potentials, and point out the differences.

For a field $\psi(\mathbf{x})$ corresponding to the quark Matsubara mode $\omega_n$ the Euclidean Lagrangian (\ref{eq:L_QCD}) reads
\begin{equation}
 \mathcal{L}_\mathrm{q} = \bar{\psi}\left[ i\gamma_0 \omega_n +\gamma_0 \mu -ig\gamma_0 A_0 +\gamma_k D_k 
	+\gamma_3 D_3\right]\psi,
\label{eq:fermion4d}
\end{equation}
where $k=1,2$ and $A_0$ is the gluonic zero mode. Note that the interaction vertex with $A_0$ does not mix the different fermionic modes, so we have separate terms like Eq.~(\ref{eq:fermion4d}) for each mode. We have used the rotational invariance to measure the correlations always in the $x_3$-direction, and therefore separate the $D_3\psi$ term from the transverse ones.

In the nonrelativistic region we should be able to treat the particle and the antiparticle separately. We choose a nonstandard representation for the Euclidean Dirac matrices,
\begin{equation}
 \gamma_0 = \left( \begin{array}{cc} 0 & 1_2 \\ 1_2 & 0 \end{array} \right), \qquad
 \gamma_i = \left( \begin{array}{cc} \epsilon_{ij}\sigma_j & 0 \\ 0 & -\epsilon_{ij}\sigma_j \end{array} \right), \qquad
 \gamma_3 = \left( \begin{array}{cc} 0 & -i \\ i & 0 \end{array} \right),
\label{eq:dirac_matrices}
\end{equation}
so that
\begin{equation}
 \gamma_0 \slashed{p} = \left( \begin{array}{cc} (p_0+ip_3)1_2 & -\epsilon_{ij}p_i \sigma_j \\
                                \epsilon_{ij}p_i \sigma_j & (p_0-ip_3)1_2
                               \end{array} \right),
\end{equation}
where $\epsilon_{ij}$ is antisymmetric and $\epsilon_{12}=+1$. This is diagonal in $p_0$ and $p_3$, so writing
\begin{equation}
 \psi = \left( \begin{array}{c} \chi \\ \phi \end{array} \right)
\end{equation}
gives us
\begin{equation}
 \mathcal{L}_\mathrm{q} = i\chi^\dagger\left( p_0 -gA_0 +D_3 \right)\chi +i\phi^\dagger\left( p_0 -gA_0 -D_3 \right)\phi
	+\phi^\dagger \epsilon_{ij}D_i\sigma_j\chi -\chi^\dagger \epsilon_{ij}D_i\sigma_j\phi,
\end{equation}
where $p_0$ again includes the chemical potential, $p_0=\omega_n-i\mu$.

As stated above, we are expanding around the screening poles $p_3 = \pm i p_0$, where one component is light while the other one is heavy compared with the dynamical scales. Working at the tree level, we can then solve the equations of motion for the heavy component and expand in $1/p_0$, getting
\begin{eqnarray}
 \mathcal{L}_\mathrm{q} &\approx & i\chi^\dagger\left[ p_0 -gA_0 +D_3 -\frac{1}{2p_0}\left( D_\bot^2 
	+\frac{g}{4i} [\sigma_i, \sigma_j] F_{ij} \right) \right] \chi \nonumber \\
 && +i\phi^\dagger\left[ p_0 -gA_0 -D_3 -\frac{1}{2p_0}\left( D_\bot^2 + \frac{g}{4i}[\sigma_i,\sigma_j]F_{ij}\right)
	\right] \phi +\mathcal{O}\left( \frac{1}{p_0^2} \right).
\label{eq:treelevelL}
\end{eqnarray}

At classical level the effective Lagrangian for quark Matsubara modes then consists of two nonrelativistic particles coupled to gluons and to a massive adjoint scalar, all in three dimensions. This can be rotated to a (2+1)-dimensional Minkowski space as in \cite{Huang:1996tz} to have a form similar to NRQCD, but for the screening states this is an unnecessary complication, and we choose to stay in Euclidean space instead.

\subsection{Matching at $g^2$ level}
\label{subsec:matching}

When going beyond the tree level, the parameters in Eq.~(\ref{eq:treelevelL}) are subject to quantum corrections. We also need to consider all other possible operators allowed by symmetries but not present at the classical level. To keep the number of possibilities finite, it is necessary to set up a power counting for the possible operators.

By definition, the effective theory contains quarks with $|\mathbf{p}_\bot| \lesssim gT$ and the off-shellness $\Delta p_3 = p_3\pm ip_0$ of the same order. Requiring the action to be of order unity, we have
\begin{eqnarray}
 \int \dd x_3\, \dd^2 \xbot \chi^\dagger \partial_3 \chi \sim 1, & \Rightarrow & \chi \sim 1/|\xbot| \sim gT \\
 \int \dd x_3\, \dd^2 \xbot A \partial^2_3 A \sim 1 & \Rightarrow & A \sim (x_3/\xbot^2)^{1/2} \sim g^{1/2} T^{1/2}\, .
\end{eqnarray}
For relativistic on-shell gluons the energy and momentum are of the same order, $p_3 \sim \mathbf{p}_\bot$. The quarks, on the other hand, are nonrelativistic, so their kinetic energy ($p_3$ in the 2+1-dimensional theory) is proportional to the momentum squared. As can be verified from the poles in quark propagators Eqs.~(\ref{eq:chipropagator}),(\ref{eq:phipropagator}), for nearly on-shell quarks in $\textrm{NRQCD}_3$, with transverse momentum $\mathbf{p}_\bot \lesssim gT$, the off-shellness is
\begin{equation}
 \Delta p_3 \sim \mathbf{p}_\bot^2/p_0 \sim g^2 T \quad \Rightarrow \quad \partial_3 \sim g^2 T \qquad \textrm{acting on quarks}.
\end{equation}

These estimates enable us to limit the number of possible operators. For example, any four-quark operator would be of the order $g^4 T^4$, while the leading order behaves as $g^2 T^3$. This means that in order to produce at most $\mathcal{O}(g^2)$ corrections the coefficient would have to be of order $g^0$, but this term is not present at tree level and can therefore be excluded. Likewise, there is a known kinematic correction $D_\bot^4/8p_0^3$ coming from the correct normalization of the spinors, but this is also of higher order ($g^4$) than we need in our calculation.

The same power counting also shows that $\gE A \sim g^{3/2} T$, which is of higher order than the derivative term $\partial_i$ in the transversal covariant derivative. Since $\partial_\bot^2$ is already of order $g^2$, we can leave out the transverse gluons. Similarly, the only term that needs matching beyond tree-level is the zero-point energy $p_0$, which we will denote by $M$ from now on. To order $g^2$ it is then sufficient to use the fermionic Lagrangian
\begin{equation}
 \mathcal{L}_\mathrm{eff}^\mathrm{f} = i\chi^\dagger\left( M -\gE A_0 +D_3 -\frac{\nabla_\bot^2}{2p_0} \right)\chi 
	+i\phi^\dagger\left( M -\gE A_0 -D_3 -\frac{\nabla_\bot^2}{2p_0} \right)\phi.
\label{eq:final_nrqcd}
\end{equation}

The one-loop correction to the parameter $M$ can be computed by matching Green's functions computed in QCD and $\textrm{NRQCD}_3$. To avoid the factors of $g^2$ arising from the normalization of the fields, we choose to match the location of the pole in quark propagator in the $p_3$-plane. In both the effective theory and the full QCD this is a gauge invariant quantity, which corresponds to the quark pole mass in the $(2+1)$-dimensional theory.

We will use the gluon propagator with an infrared regulator $\lambda$ as well as a gauge parameter $\xi$ and use dimensional regularization throughout to control the ultraviolet behavior. We will see that in the final results the dependence on $\xi$ vanishes and that we are allowed to take the limit $\lambda\to 0$, $\epsilon \to 0$. The gluon propagator in full QCD reads
\begin{equation}
 \langle A_\mu^a(p) A_\nu^b(q)\rangle = \delta^{ab}(2\pi)^{4-2\epsilon} \delta^{(4-2\epsilon)}(p+q)\left[ \left(
	\delta_{\mu \nu}-\frac{p_\mu p_\nu}{p^2}\right)\frac{1}{p^2+\lambda^2} + \frac{p_\mu p_\nu}{p^2}\frac{\xi}{p^2+\xi\lambda^2} \right].
\end{equation}

\FIGURE[t]{
\parbox{0.9\textwidth}{
\begin{center}
\begin{fmfgraph*}(80,30)
 \fmfset{arrow_len}{3mm}
 \fmfleft{l}
 \fmfright{r}
 \fmf{fermion,label=$p$}{l,v1}
 \fmf{fermion,label=$p-q$,tension=0.5}{v1,v2}
 \fmf{fermion}{v2,r}
 \fmf{photon,left,tension=0,label=$q$,label.side=right}{v1,v2}
\end{fmfgraph*}
\end{center} }
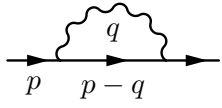
\caption{One-loop quark self-energy correction.}
\label{fig:quark_se} }

The one-loop self-energy correction to the quark propagator is given by Fig.~\ref{fig:quark_se},
\begin{equation}
 \Sigma(p) = i g^2 \CF \sumint_q \frac{\gamma_\mu(\slashed{p}-\slashed{q})\gamma_\mu}{(p-q)^2(q+\lambda^2)}
	-ig^2 \CF \sumint_q \frac{\slashed{q}(\slashed{p}-\slashed{q})\slashed{q}}{q^2(p-q)^2}\left( 
	\frac{1}{q^2+\lambda^2} -\frac{\xi}{q^2+\xi\lambda^2} \right),
\label{eq:quark_se}
\end{equation}
where we have used the convenient short-hand notation
\begin{equation}
 \sumint_q \equiv \sum_{q_0=\omega_n^b} \left( \frac{\Lambda^2 e^\gamma}{4\pi}\right)^\epsilon \int \! \frac{\dd^d q}{(2\pi)^d}
\end{equation}
for dimensionally regularized sum-integrals in the $\MSbar$ scheme.

The combination $i\slashed{p}-\Sigma(p)$ should annihilate the on-shell spinor $u(\mathbf{p})$ order by order in perturbation theory. When computing the order $g^2$ correction $\Sigma(p)$, we can use the leading order results $p^2=0$ and $\slashed{p}u(\mathbf{p})=0$ to drop terms proportional to $p^2$ and $\slashed{p}$, since $\Sigma(p)$ is already multiplied by $g^2$. The longitudinal part then vanishes and the result is independent of the gauge parameter $\xi$. The remaining expression is still rotationally invariant, but we can choose to match the Green's functions at some specific momentum and choose $\mathbf{p}_\bot=0$. When multiplied from the left by $\gamma_0$, the expression becomes diagonal in the representation of Dirac matrices given by Eq.~(\ref{eq:dirac_matrices}), and we can concentrate on, say, $[\gamma_0 \Sigma]_{11}$, which is
\begin{equation}
 [\gamma_0 \Sigma]_{11} = -i(D-2) g^2 \CF \sumint_q \frac{p_0+ip_3 -q_0-iq_3}{(p-q)^2(q^2+\lambda^2)}.
\end{equation}
We again use the tree-level relation $p_3=ip_0$ inside the integral, which then reduces to a sum of a bosonic and a fermionic tadpole, both infrared finite, giving the pole position
\begin{equation}
 p_3 \approx i\left[ p_0 +g^2\CF \frac{T^2}{8p_0}\left(1+\frac{\mu^2}{\pi^2 T^2}\right)\right].
\end{equation}

The $\mathrm{NRQCD}_3$ Feynman rules can be seen from Lagrangian in Eq.~(\ref{eq:final_nrqcd}). In particular, the propagators are
\begin{eqnarray}
 \langle \chi_u(p)\chi^*_v(q) \rangle &=& \delta_{uv}(2\pi)^3\delta(p-q) \frac{-i}{M+ip_3+\mathbf{p}_\bot^2/2p_0}
	\label{eq:chipropagator} \\
 \langle \phi_u(p)\phi^*_v(q) \rangle &=& \delta_{uv}(2\pi)^3\delta(p-q) \frac{-i}{M-ip_3+\mathbf{p}_\bot^2/2p_0}
	\label{eq:phipropagator}
\end{eqnarray}
or, in the configuration space,
\begin{eqnarray}
 \langle \chi_u(x)\chi^*_v(y) \rangle &=& -i\delta_{uv}\theta( \omega_n(x_3-y_3)) \frac{p_0}{2\pi(x_3-y_3) }
	e^{-M(x_3-y_3)-\frac{p_0(\xbot -\mathbf{y}_\bot)}{2(x_3-y_3)} } \\
 \langle \phi_u(x)\phi^*_v(y) \rangle &=& -i\delta_{uv}\theta( \omega_n(y_3-x_3)) \frac{p_0}{2\pi(y_3-x_3) }
	e^{-M(y_3-x_3)-\frac{p_0(\xbot -\mathbf{y}_\bot)}{2(y_3-x_3)} }.
\end{eqnarray}
These equations show clearly that for $\omega_n > 0$ the $\chi$ field propagates forward and $\phi$ backward in $x_3$, the time coordinate in the (2+1) -dimensional theory. For $\omega_n < 0$ the roles are reversed.

In \cite{Laine:2003bd} we expanded the $\mathrm{NRQCD}_3$ propagators in $1/p_0$, treating the kinetic term as a perturbation. This works well in the case of a single heavy quark, but, as discussed in \cite{Luke:1999kz}, will run into problems in NRQCD at two-loop level. We will in this paper follow \cite{Luke:1999kz} in keeping the $\mathbf{p}_\bot^2/2p_0$ term resummed into the propagator. This also means that we have to use the multipole expansion, which in practice means that the gluons are not allowed to transfer transverse momentum to quarks. As a consistency check, we have shown that this formulation of the effective theory is able to reproduce exactly the free theory result Eq.~(\ref{eq:vapaaskalaari}).

On the effective theory side using the different propagators and adding chemical potentials does not change the result of \cite{Laine:2003bd}; the one-loop correction to quark self-energy vanishes at the pole in $\mathrm{NRQCD}_3$. The pole is then located at $p_3 \approx iM$ and we can match these values to give
\begin{equation}
 	M = p_0 +g^2\CF \frac{T^2}{8p_0}\left(1+\frac{\mu^2}{\pi^2 T^2}\right) =
	\omega_n -i\mu +g^2\CF \frac{T^2}{8(\omega_n-i\mu)}\left(1+\frac{\mu^2}{\pi^2 T^2}\right)
\label{eq:matched_M}
\end{equation}
The above equation was derived around the pole $p_3=ip_0$ and for the (11)-component of $\Sigma(p)$, so it gives the mass of the $\chi$ field. A similar computation for the pole $p_3=-ip_0$ shows that the $\phi$ field has the same mass. We made no assumptions about the sign of $\omega_n$ here, and hence Eq.~(\ref{eq:matched_M}) is valid for all Matsubara modes. However, as we will see below, for negative modes it is $-M$ that enters the equations as a mass paramater. It should be noted that for the lightest modes $\omega_n = \pm \pi T$, which eventually are the only important ones here, the real part of the mass is independent of $\mu$.


\section{Solution for the screening states}
\label{sec:solution}

With the effective Lagrangian of Eqs.~(\ref{eq:bosonicL}), (\ref{eq:final_nrqcd}) at hand, we are now ready to compute the correlators. The various bilinears in Eq.~(\ref{eq:general_bilinear}) can be written as products of the $\chi$ and $\phi$ fields with $\sigma^i$ and flavor matrices between them. The details can be found in \cite{Laine:2003bd}. Since the propagators (\ref{eq:chipropagator}), (\ref{eq:phipropagator}) as well as the couplings in Eq.~(\ref{eq:final_nrqcd}) are diagonal in spin indices, and $(\sigma^i)^2=1$, the spin structure does not affect the result. The flavor structure, on the other hand, will be significant if we allow for different chemical potential for each quark flavor.

As discussed in the beginning of Sec.~\ref{sec:et}, and verified by the power counting rules above, we have to take into account all diagrams of type Fig.~\ref{fig:diagram_types}(b,c) with an arbitrary number of soft gluons. The matching computation already took care of the hard ($\omega_n^b\neq 0$) part of Fig.~\ref{fig:diagram_types}(c), and the non-static modes in Fig.~\ref{fig:diagram_types}(b) do not contribute to the pole location. In the nonrelativistic theory we can take the gluon exchange to be instantaneous, and compute the static potential for the $\phi^*\chi$ pair by integrating out the gauge fields. The screening states can then be solved from the familiar Schr\"odinger equation.

It should be noted here that it is enough to compute the potential to order $\gE^2$. To see this, notice that the general structure of the potential can be written as an expansion in $\gE^2 r$,
\begin{equation}
 V(r) \sim \gE^2 \ln r + \gE^4 r + \mathcal{O}(\gE^6 r^2).
\end{equation}
The Schr\"odinger equation gives to lowest order
\begin{equation}
 \frac{1}{p_0}\frac{\partial^2}{\partial r^2} \sim V(r) \sim \gE^2 \ln r,
\end{equation}
so $1/r \sim \sqrt{p_0 \gE^2} \sim gT$, or $\gE^2 r \sim g$. The leading term $\gE^2 \ln r$ coming from the one-gluon exchange is then sufficient for finding the $g^2$ correction to the screening mass. It should be noted that this term already gives a confining potential, so there is no qualitative difference in leaving out the linear term. There might still be infrared divergences when computing the scale inside the logarithm, but as we will see, this term is completely finite in $\mathrm{NRQCD}_3$ and no further resummations are needed.

The potential can be found by introducing a point-splitting in the correlator to put the quarks a finite distance $\mathbf{r}$ apart, then finding the Schr\"odinger equation satisfied by the correlator and in the end letting $\mathbf{r} \to 0$. This calculation for $\mu=0$ was carried out in \cite{Laine:2003bd}, and adding the chemical potentials and using propagators with the transverse kinetic terms included does not change the result, which is
\begin{equation}
 V(\mathbf{r}) = \frac{\gE^2 \CF}{2\pi}\left( \ln\frac{\mE r}{2} +\gamma_E -K_0(\mE r) \right).
\label{eq:potential}
\end{equation}
Since the computation of the potential was not shown explicitly in \cite{Laine:2003bd}, we give the details in App.~\ref{app:potential}.

The correlator $C(\mathbf{r},z)$ for a $\phi^*_\mathrm{i}\chi_\mathrm{j}$ pair, where i,j are flavor indices, then satisfies the equation
\begin{equation}
 \left[ \partial_z \pm(M_\mathrm{i} +M_\mathrm{j}) -\frac{1}{\pm 2\bar{p}_\mathrm{0ij}}\nabla^2_\mathbf{r} 
	+V(\mathbf{r}) \right]C(\mathbf{r},z) = 2N_c \,\delta(z)\delta^2(\mathbf{r}),
\label{eq:corr_schrod}
\end{equation}
where $\bar{p}_\mathrm{0ij}$ is the reduced mass of the system,
\begin{equation}
 \frac{1}{\pnij} \equiv  \frac{1}{\omega_n-i\mu_\mathrm{i}} + \frac{1}{\omega_n-i\mu_\mathrm{j}}
\label{eq:pnij_def}
\end{equation}
and the $+$ and $-$ signs stand for positive and negative Matsubara modes $\omega_n=\pm \pi T$, respectively. The exponential falloff is determined by the eigenvalues of the Schr\"odinger equation
\begin{equation}
 \left[ \pm(M_\mathrm{i} +M_\mathrm{j}) -\frac{1}{\pm 2\bar{p}_\mathrm{0ij}}\nabla^2_\mathbf{r} 
	+ V(\mathbf{r}) \right]\Psi_0 = m_\mathrm{full}\Psi_0,
\label{eq:schrodinger}
\end{equation}
with $\Psi_0$ some ground state wave function. It should be noted that the parameters of the modes $\pm \pi T$ are related by $M_- = -M_+^*$ and $p_{0-}=-p_{0+}^*$, and that the potential is real, so the complex conjugate of Eq.~(\ref{eq:schrodinger}) gives
\begin{equation}
 m_{\mathrm{full},-} = m_{\mathrm{full},+}^*.
\label{eq:massrelation}
\end{equation}
Assuming the ground state is unique (up to normalization), the full correlator of Eq.~(\ref{eq:pointplanedef}) which is the sum over both positive and negative modes, behaves as
\begin{eqnarray}
 \lefteqn{ C_z[O^a, O^b] \approx } \nonumber \\ 
 & \approx & \int \! \dd^2 \mathbf{R} \,
	\langle \phi^*_+(\mathbf{R},z)F^a\chi_+(\mathbf{R},z) \chi^*_+(0)F^b\phi_+(0) \rangle +
	\langle \chi^*_-(\mathbf{R},z)F^a\phi_-(\mathbf{R},z) \phi^*_-(0)F^b\chi_-(0) \rangle \nonumber \\
 &\propto & \sum_\mathrm{ij} F^a_\mathrm{ij}F^b_\mathrm{ji}\left[ e^{i\alpha_\mathrm{ij}} \exp(-m_\mathrm{full,ij}z) +e^{-i\alpha_\mathrm{ij}} \exp(-m^*_\mathrm{full,ij}z) \right] \nonumber \\
 &=& \sum_\mathrm{ij} F^a_\mathrm{ij}F^b_\mathrm{ji}\, 2\cos[\mathrm{Im}(m_\mathrm{full,ij})z -\alpha_\mathrm{ij}]\, \exp[-\mathrm{Re}(m_\mathrm{full,ij})z ],
\end{eqnarray}
where $\alpha_\mathrm{ij}$ is an unimportant overall phase of the $\phi_\mathrm{i}^*\chi_\mathrm{j}$ correlator. The correlator is evidently real as it should be and behaves as $\exp(-mz)\cos(\mu z)$, compatible with the leading order term in Eq.~(\ref{eq:vapaayleinen_appr}). In particular, the screening mass is given by the real part $\mathrm{Re}(m_\mathrm{full})$, while the imaginary part contributes to the oscillation with $z$. Because of the relation Eq.~(\ref{eq:massrelation}), we can restrict the following analysis to the fields corresponding to the positive Matsubara mode $+\pi T$.

The eigenvalues of the Schr\"odinger equation (\ref{eq:schrodinger}) have to be found numerically. We change to dimensionless variables
\begin{equation}
 \hat{r} \equiv \mE r, \qquad \gE^2\frac{\CF}{2\pi}\hat{E}_0 \equiv m_\mathrm{full} -M_\mathrm{i} -M_\mathrm{j}, \qquad
	\rho \equiv \frac{\gE^2\CF}{2\pi}\frac{2\bar{p}_\mathrm{0ij}}{\mE^2}
\label{eq:dimless_params}
\end{equation}
giving for a spherically symmetric state
\begin{equation}
\left[ -\left( \frac{\dd^2}{\dd \hat{r}^2} +\frac{1}{\hat{r}}\frac{\dd}{\dd\hat{r}}\right) 
	+\rho \left( \ln \frac{\hat{r}}{2} +\gamma_E -K_0(\hat{r}) -\hat{E}_0\right) \right]\Psi_0 =0.
\label{eq:dimless_schrod}
\end{equation}

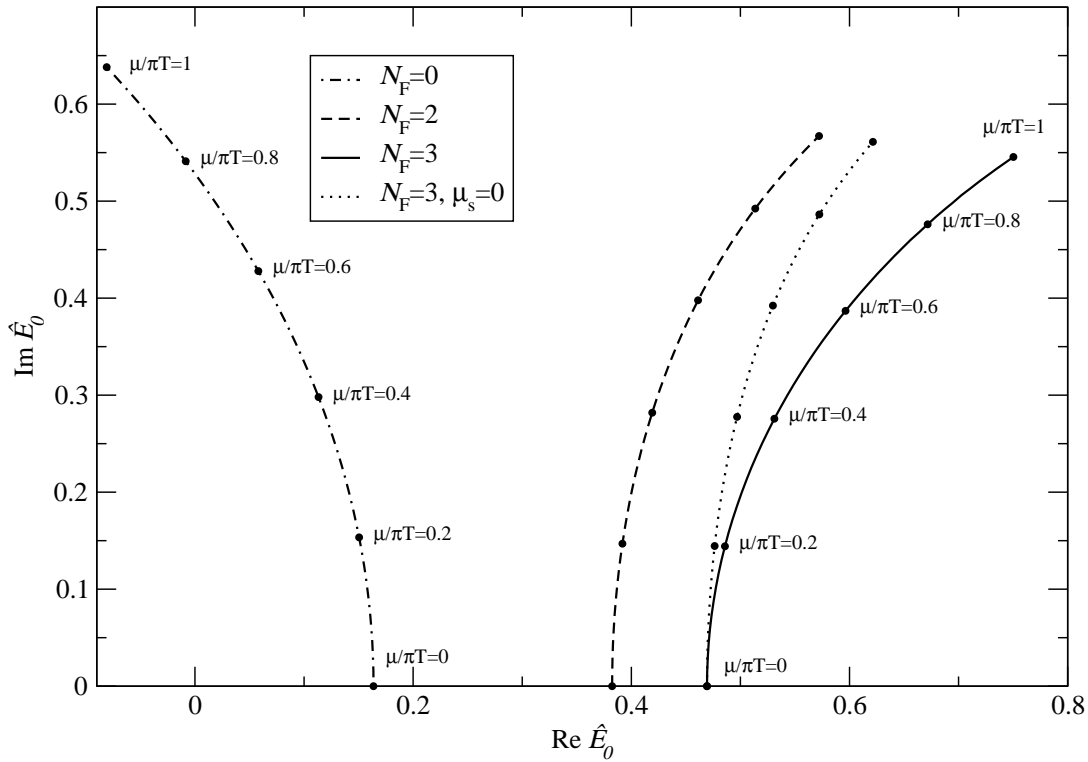
\EPSFIGURE[t]{poles.eps,width=0.95\textwidth}{The $\hat{E}_0$ eigenvalue
with the lowest real part for $\mu_S/\pi T=0\ldots1$.
\label{fig:scalar_poles} }

The allowed values for $\hat{E}_0$ are found by integrating Eq.~(\ref{eq:dimless_schrod}) out from small $\hat{r}$ and requiring square integrability. This is somewhat more complicated than in the $\mu=0$ case, since now both $\rho$ and $\hat{E}_0$ are complex, so Eq.~(\ref{eq:dimless_schrod}) actually represents two coupled equations for $\mathrm{Re}(\Psi_0)$ and $\mathrm{Im}(\Psi_0)$. Modified versions of the integration routines we used in \cite{Laine:2003bd} can be applied, but the values of $\hat{E}_0$ now have to be searched in the complex plane instead of the real axis. In addition, the result depends on the chemical potentials in a nontrivial way, and the numerical computation has to be carried out separately for each $\mu/\pi T$.

\subsection{Results and discussion}

For isoscalar chemical potential ($\mu_\mathrm{u}=\mu_\mathrm{d} \equiv \mu_S$) the result of solving Eq.~(\ref{eq:dimless_schrod}) numerically is plotted in Fig.~\ref{fig:scalar_poles}. Shown is the location of the pole in the complex $\hat{E}_0$-plane for $0\leq \mu_S/\pi T \leq 1$. For negative values of $\mu_S$ Eq.~(\ref{eq:dimless_schrod}) shows that the solution is $\hat{E}_0(-\mu)=\hat{E}_0^*(\mu)$. As can be seen from the figure, the pole moves on a quadratic curve off the real axis and towards larger $\mathrm{Re}(\hat{E}_0)$, except for the $N_F=0$ result. In the $N_F=3$ case we also study the situation where only two of the quarks have finite chemical potentials, whereas the density of the third quark vanishes. This should correspond to the situation in the heavy ion collisions, where the two incoming nuclei have no net strangeness, but the strange quark mass is small compared to the temperature. For isovector chemical potential the parameter $\rho$ above is real, so the eigenvalues stay on the real axis.

The screening mass depends on the real part of $\hat{E}_0$ through
\begin{equation}
 \mathrm{Re}(m_\mathrm{full}) = \mathrm{Re}(M_\mathrm{i}+M_\mathrm{j}) +\gE^2\frac{\CF}{2\pi}\mathrm{Re}(\hat{E}_0)
	= 2\pi T +g^2 T \frac{\CF}{2\pi}\left( \half + \mathrm{Re}(\hat{E}_0) \right),
\label{eq:fullmass}
\end{equation}
where only $\hat{E}_0$ depends on the chemical potentials. In Fig.~\ref{fig:real_part} we plot $\mathrm{Re}(\hat{E}_0)$ vs.~$\mu$ for both isoscalar and isovector chemical potentials.
In the isoscalar case the screening mass increases slowly with $\mu_S$ for dynamical fermions, while there is a decrease in the $N_F=0$ (quenched quarks) case. The derivative of the real part with respect to $\mu$ vanishes at $\mu=0$ for all chemical potentials, which follows directly from the symmetry $\hat{E}_0(-\mu)=\hat{E}_0^*(\mu)$.
\EPSFIGURE[bt]{real_vs_mu.eps,width=0.95\textwidth}{
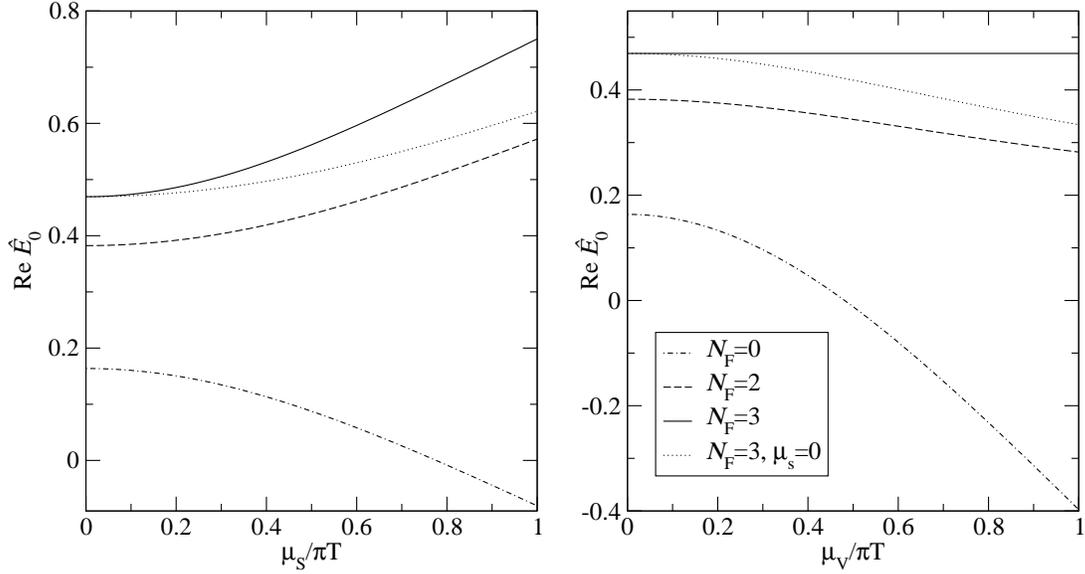
The real part of $\hat{E}_0$, for isoscalar (left) and isovector (right) 
chemical potential. \label{fig:real_part}}

In the case of isovector chemical potential the mass decreases as a function of $\mu_V$ when $N_F\leq 2$. At $N_F=3$ the $\mu_V$-dependence in $\bar{p}_0 \propto (1+\hat{\mu}^2)$ and $\mE^2 \propto (6+N_F+3N_F\hat{\mu}^2)$ cancels exactly in $\rho$, leading to a $\mu_V$-independent mass, but this is a plain numerical coincidence with no deeper physical significance. With even more dynamical flavors the mass would increase with $\mu_V$, but we have not done any numerics in those cases.

The numerical value of the correction is small compared to the leading order result. For $N_F=2$, Eq.~(\ref{eq:fullmass}) gives numbers in the interval
\begin{equation}
 \mathrm{Re}(m_\mathrm{full}) \approx 2\pi T +\gE^2 \times \left\{ 
	\begin{array}{lr}
	 0.227, & \mu_S/\pi T = 1.0 \\
	 0.187, & \mu/\pi T = 0.0 \\
	 0.166, & \mu_V/\pi T = 1.0 \\
	\end{array} \right.
\end{equation}
At $T\sim 2T_c$ the effective coupling $\gE$ is estimated \cite{Laine:2005ai} to be $\gE^2/T \approx 2.2$, so the first correction is about 6-8\%. However, because the gauge coupling is large, this does not guarantee that the perturbative series would converge rapidly.

To sum up, the general behavior of the screening mass in the presence of dynamical fermions is that it increases with isoscalar chemical potential and slightly decreases with the isovector one. Note, however, that the behavior of the correlator differs from a simple exponential, in particular at short distances. When the correlator is measured only at short distances, it falls off faster because of the cosine term (see Fig.~\ref{fig:vapaa}), showing an apparent increase in screening. To see this more specifically, we note that simply fitting an exponential of the form $C\exp(-m z)$, where $C$ and $m$ are the fitting parameters, to the plane-plane correlator $\propto \exp(-Mz)\cos(\mu z)$ gives
\begin{equation}
 (m-M)(m+M)^2 = \mu^2(3m+M) \quad \Rightarrow \quad m \approx M\left(1 + \frac{\mu^2}{M^2} -\frac{\mu^4}{4M^4}\right).
\label{eq:m_fit}
\end{equation}
One should therefore be careful when extracting the screening mass from the correlator, since fitting a wrong kind of a function will falsely interpret the oscillation length as part of the exponential falloff. This may also partly explain the considerably weaker $\mu$ dependence observed on lattice in the isovector case. As can be seen in Eq.~(\ref{eq:vapaayleinen_appr}), there is no oscillation at leading order when $\mu_\mathrm{u}=-\mu_\mathrm{d}$.

It is not very clear how the screening mass should be determined in the finite density case if the functional form of the correlator is not known. At $\mu=0$ one often defines the effective (distance-dependent) mass
\begin{equation}
 m(z) = -\frac{1}{C_z}\frac{\partial C_z}{\partial z},
\end{equation}
which slowly converges to the screening mass defined as $\lim_{z\to \infty} m(z)$. At finite density this obviously does not work, because $C_z$ vanishes at $z \sim 1/\mu$. We have chosen to define the mass strictly as the coefficient of the exponential falloff, while keeping in mind that the correlator is generally more complicated than in the $\mu=0$ case.

\subsection{Comparison with lattice}

At vanishing quark chemical potentials the high temperature mesonic screening masses have been measured on the lattice twenty years ago \cite{DeTar:1987ar,*DeTar:1987xb,Gottlieb:1987gz}. Recent results can be found in \cite{Wissel:2005pb,Gavai:2006fs}. They have already been discussed more thoroughly in \cite{Laine:2003bd}. The masses are found to be close to the ideal gas limit, with $\pi$ meson lying somewhat below $\rho$. The deviation from the free theory result in those lattice computations is invariably negative, with masses around 5-10\% below $2\pi T$, whereas our NLO analytic calculation gives screening masses slightly above the noninteracting result. The results of the current paper behave likewise, since the $\mu=0$ limit of our result agrees with our previous paper.

Lattice measurements of hadron screening masses at finite density have been performed only recently, mostly because of the complex action which makes the simulations difficult. In \cite{Pushkina:2004wa} the first two coefficients in the Taylor expansion of the mass around $\mu=0$ are measured for both isoscalar and isovector chemical potentials. The simulations are done at $\mu=0$, and the result is of course valid only when the chemical potential is small. The leading term is just the screening mass at $\mu=0$, and again this lies a little below the free theory value, $\pi$ (pseudoscalar) below $\rho$ (vector). The first derivatives vanish at $\mu=0$; for $\mu_S$ this follows from a symmetry, and for $\mu_V$ it is measured to be consistent with zero. This agrees well with our results.

The $\mu$-dependence of the screening masses is given by second derivatives measured in \cite{Pushkina:2004wa}. They observe that for both $\pi$ and $\rho$ the second order response increases with temperature in the isoscalar case, while in the isovector case it approaches zero from below. Note that the simulations in \cite{Pushkina:2004wa} were done with two dynamical quark flavors, corresponding to dashed lines in Fig.~\ref{fig:real_part}. The isoscalar case qualitatively agrees with our result, with the mass increasing roughly quadratically near $\mu_S=0$. Also in the isovector case the slowly decreasing mass in Fig.~\ref{fig:real_part} agrees with the small negative second derivative in \cite{Pushkina:2004wa}. It should be noted that at least in our analysis the small value of the second derivative is a consequence of the choice $N_F=2$, giving the dimensionless parameter $\rho$ in Eq.~(\ref{eq:dimless_schrod}) as
\begin{equation}
 \rho = \frac{1+\hat{\mu}^2}{1+\frac{3}{4}\hat{\mu}^2}, \qquad \hat{\mu}\equiv \frac{\mu_V}{\pi T},
\end{equation}
which depends only weakly on $\mu_V$. For $N_F=3$ the derivative vanishes completely, and $N_F=4$ would give a positive second derivative also in the isovector case.

To get some quantitative comparison, we can also try to fit a quadratic curve to our data. At larger $\mu/\pi T$ the behavior of our result is clearly something else than simple $\mu^2$, so we quite arbitrarily restrict the fitted region to $\mu/\pi T < 0.5$. Assuming that the first derivative vanishes, we look for the coefficients of  $\mathrm{Re}(\hat{E}_0) = c_1 + c_2 (\mu/\pi T)^2$, which gives the derivatives as
\begin{equation}
 \frac{\dd^2 \mathrm{Re}(\hat{E}_0)}{\dd \hat{\mu}_S^2} = \left\{
	\begin{array}{rl}
	 -0.62, & (N_F=0) \\	0.45, & (N_F=2) \\	0.75, & (N_F=3) \\ 0.34, & (N_F=3, \mu_\mathrm{s}=0)
	\end{array} \right.
\qquad
 \frac{\dd^2 \mathrm{Re}(\hat{E}_0)}{\dd \hat{\mu}_V^2} = \left\{
	\begin{array}{rl}
	 -1.42, & (N_F=0) \\	-0.31, & (N_F=2) \\	0.00, & (N_F=3) \\ -0.41, & (N_F=3, \mu_\mathrm{s}=0)
	\end{array} \right.
\end{equation}
for isoscalar and isovector, respectively. In terms of physical parameters
\begin{equation}
 T\frac{\dd^2 m}{\dd \mu^2} = \frac{\gE^2\CF}{2\pi^3 T}\frac{\dd^2\mathrm{Re}(\hat{E}_0)}{\dd \hat{\mu}^2},
\end{equation}
which depends on $\gE$. Using $\gE^2/T \approx 2.2$ as we did earlier, this gives second derivatives of order $T\dd^2 m/\dd \mu^2 \sim \pm 0.02$. For isoscalar they are much smaller than in \cite{Pushkina:2004wa}, but the difference may partly follow from the different definitions of mass. For example, including the cosine term like in Eq.~(\ref{eq:m_fit}) would contribute an additional $2T/M \approx 1/\pi$.

\section{Conclusions and outlook}
\label{sec:conclusions}

This paper has extended our previous computation of mesonic screening masses to situations where not only high temperature but also a finite chemical potential is present. For flavor non-singlet operators we have computed the full correlator to leading order in perturbation theory and derived a next-to-leading order correction to the screening mass. We have studied how the mass depends on the chemical potentials of the quarks as well as on the number of dynamical flavors.

Our method is valid for an arbitrary combination of chemical potentials. The numerical value of the NLO correction was computed for two distinct cases. We found that when $N_F>0$, the screening mass increases with $\mu$ if both quarks in the operator have the same (isoscalar) $\mu$, whereas for opposite chemical potentials the mass slowly decreases. With quenched quarks the mass decreases in both cases. The data near $\mu=0$ is well described by the quadratic curve $m(\mu)=m(0)+C\mu^2$, but the coefficients we get for the second-order response are an order of magnitude smaller than in the recent lattice data.

The meson correlator looks like a simple exponential only at asymptotic distances. At shorter scales the contributions from higher Matsubara modes as well as from factors of $1/z$ modify the behavior, as can be seen from plots of the effective mass. At finite density the situation is even more complicated, because the chemical potentials manifest themselves as oscillatory terms, which can have a large effect on the short distance behavior. In fact, at scales where the correlator is significantly different from zero the screening mass alone does not give a sufficient description. It has its virtue in being a simple parameter characteristic of screening phenomena, accessible both analytically and on the lattice where there is a long history of measuring the screening correlator. However, it is not easy to measure the correlator on the lattice at large distances because of the exponential vanishing of the correlator and large lattices needed. On the other hand, to extract the mass from short-distance data we need to know the general behavior of the correlator.

We have not touched the issue of flavor singlets in this paper. At large distance their correlators are dominated by couplings to different glueball states, whose masses are lower than $2\pi T$. We have worked out the couplings in \cite{Laine:2003bd}, and the glueball masses have been measured in \cite{Hart:2000ha} also for finite chemical potentials. One notable change in finite density is that the $\Tr A_0^3$ operator corresponding to $\bar{\psi}\gamma_0\psi$ now couples to $\Tr A_0^2$, which has a much lower mass.

Even the finite chemical potential does not change the fact that our NLO correction to the screening mass is strictly positive, unlike all the lattice measurements. At asymptotically high temperature the perturbation theory should give correct results, so we expect the screening mass to go above $2\pi T$ as temperature is increased. On the other hand, implementing our theory on the lattice could show if the higher order and nonperturbative terms bring the result below the free theory value at temperatures close to the phase transition. The computed correction is small even when $g$ is large, but the next linear $\mathcal{O}(g^3)$ term in the potential could easily be of the same magnitude. Our theory could also be used to extract other features of the correlators besides masses. This would require matching also the fields to $g^2$ order and probably solving the Schr\"odinger equation in some approximation.

\acknowledgments

This work was supported by the Jenny and Antti Wihuri Foundation and the V\"ais\"al\"a Foundation. I would like to thank K.~Kajantie and M.~Laine for useful discussions and comments.


\appendix
\section{Correlators in free theory}
\label{app:freetheory}

\subsection{Isoscalar chemical potential}

For isoscalar chemical potential $\mu_\mathrm{u} = \mu_\mathrm{d} \equiv \mu$, or for flavor singlets the correlator $C_z[O^a, O^b]$ in Eq.~(\ref{eq:pointplanedef}) can be calculated explicitly in free theory. The details depend somewhat on the spin structure of the operator, we give here the computation for the scalar (and pseudoscalar, up to a sign):
\begin{eqnarray*}
\lefteqn{ C_z[S^a, S^b] = \int_0^{1/T}\!\!\dd\tau \int\!\dd^2\xbot\, \langle \bar{\psi}(\tau,\xbot,z) F^a \psi(\tau,\xbot,z)
  \bar{\psi}(0,\mathbf{0}) F^b \psi(0,\mathbf{0})\rangle } \\
 &=& -\half \delta^{ab} \int_0^{1/T}\!\!\dd\tau \int\!\dd^2\xbot\, \langle \psi(\tau,\xbot,z)\bar{\psi}(0,\mathbf{0})\rangle
	\langle \psi(0,\mathbf{0})\bar{\psi}(\tau,\xbot,z)\rangle \\
 &=& \frac{N_c}{2} \delta^{ab} \int_0^{1/T}\!\!\dd\tau \int\!\dd^2\xbot\, T^2 \sum_{mn} \int_{pq} 
	\frac{\Tr\, \slashed{p}\slashed{q}}{p^2 q^2} e^{i(\omega_m^f-\omega_n^f)\tau +i(p_\bot-q_\bot)\cdot \xbot +i(p_3-q_3)z} \\
 &=&  2N_c T \delta^{ab} \sum_n \int_p \int_{-\infty}^\infty \frac{\dd q_3}{2\pi}
	\frac{(\omega_n^f-i\mu)^2+p_\bot^2 +p_3 q_3}{ [(\omega_n^f-i\mu)^2+p_\bot^2+p_3^2] [(\omega_n^f-i\mu)^2+p_\bot^2+q_3^2]}
	e^{i(p_3-q_3)z}.
\end{eqnarray*}
The chemical potential and Matsubara frequencies only appear in terms like $(\omega_n^f -i\mu)^2$, so we can restrict the summation to positive modes by adding the term with $\mu \to -\mu$. The poles in $p_3$ and $q_3$ can be found at
\begin{equation}
 p_3,q_3 = \pm i \sqrt{ (\omega_n^f -i\mu)^2 + p_\bot^2 } \equiv \pm i \omega_n(p_\bot),
\end{equation}
where the square root is well defined off the negative real axis, and we choose the branch with positive real part. Integration over
$p_3$ and $q_3$ then gives
\begin{eqnarray}
 \lefteqn{ C_z[S^a, S^b] = \frac{N_cT}{2} \delta^{ab}\sum_{n>0} \int\! \frac{\dd^2 p}{(2\pi)^2} 
	\frac{(\omega_n^f-i\mu)^2 +p^2 +\omega_n(p)^2 }{\omega_n(p)^2} e^{-2 \omega_n(p) z} + (\mu \to -\mu) } \nonumber \\
 &=& N_c T\delta^{ab} \sum_{n>0} \int\! \frac{\dd^2 p}{(2\pi)^2} e^{-2 \omega_n(p) z} + (\mu \to -\mu) 
 = \frac{N_c T}{2\pi}\delta^{ab} \sum_{n>0} \int_{\omega_n^f-i\mu}^\infty \! \dd \omega\, \omega e^{-2\omega z} + (\mu \to -\mu) \nonumber \\
 &=& \frac{N_cT}{2\pi}\delta^{ab} \sum_{n>0} \left( -\frac{\dd}{\dd z}\right) \frac{1}{4z} e^{-2(\omega_n^f-i\mu) z} + (\mu \to -\mu) \nonumber \\
 &=& -\delta^{ab}\frac{\dd}{\dd z} \frac{N_c T}{8\pi z} \frac{1}{2\sinh 2\pi T z} e^{2i\mu z} + (\mu \to -\mu) 
 = -\delta^{ab} \frac{\dd}{\dd z} \frac{N_c T}{8\pi z} \frac{\cos 2\mu z}{\sinh 2\pi T z} \nonumber \\
 &=& \delta^{ab} \frac{N_c T}{8\pi z \sinh 2\pi T z}\left( 2\pi T \coth 2\pi Tz \, \cos 2\mu z + 2\mu \sin 2\mu z +\frac{1}{z} \cos 2\mu z \right).
\end{eqnarray}

\subsection{Arbitrary chemical potential}

When no restrictions are imposed on the chemical potentials appearing in the correlator of Eq.~(\ref{eq:pointplanedef}), even the free theory result becomes very complicated. Proceeding in the same way as we did above leads to
{\setlength\arraycolsep{2pt}
\begin{eqnarray*}
 C_z[S^a, S^b] &=& F^a_\mathrm{ij}F^b_\mathrm{ji} \frac{N_c T^2}{4\pi}\left( e^{-\Delta\mu_\mathrm{ij}/T} -1 \right)\times \nonumber \\
	&&\times \sum_{m,n} \left[ \frac{1}{\Delta\mu_\mathrm{ij}+i2\pi T(m-n)} \frac{\dd}{\dd z} \frac{1}{z} 
	e^{-(\sqrt{(\omega_m-i\mu_\mathrm{i})^2} +\sqrt{(\omega_n-i\mu_\mathrm{j})^2})z} \right. \nonumber \\
	&&\phantom{\times\sum} -\left. (\Delta\mu_\mathrm{ij}+i2\pi T(m-n))\int_z^\infty \!\dd z' \frac{1}{z'} 
	e^{-(\sqrt{(\omega_m-i\mu_\mathrm{i})^2} +\sqrt{(\omega_n-i\mu_\mathrm{j})^2})z'} \right],
\end{eqnarray*}}
where $\Delta\mu_\mathrm{ij}\equiv \mu_\mathrm{i}-\mu_\mathrm{j}$. We will also use the shorthand notation $\bar{\mu}_\mathrm{ij} \equiv \mu_\mathrm{i}+\mu_\mathrm{j}$ below.

The dominant contribution to the Matsubara sum comes from the lightest modes $\omega = \pm \pi T$, with the remaining sum suppressed by $\exp(-2\pi T z)$.  Summing only over the first modes is then a good approximation to the whole expression. We drop those subleading terms and get
\begin{eqnarray}
 \lefteqn{ C_z[S^a, S^b] = F^a_\mathrm{ij}F^b_\mathrm{ji} \frac{N_cT^2}{2\pi} \left( e^{-\Delta\mu_\mathrm{ij}/T}-1\right) \left\{
	\frac{\dd}{\dd z} \frac{1}{z} \left[ \frac{1}{\Delta\mu_\mathrm{ij}} \cos \bar{\mu}_\mathrm{ij}z \right. \right. } \nonumber \\ 
 &&{}+\left. \frac{1}{(\Delta\mu_\mathrm{ij})^2+(2\pi T)^2}( \Delta\mu_\mathrm{ij}\cos\Delta\mu_\mathrm{ij}z 
	+2\pi T \sin \Delta\mu_\mathrm{ij}z )\right] e^{-2\pi Tz} \nonumber \\
 &&{}- \left. \int_z^\infty \! \frac{\dd z'}{z'}\left[ \Delta\mu_\mathrm{ij}\cos \bar{\mu}_\mathrm{ij}z' 
	+\Delta\mu_\mathrm{ij} \cos\Delta\mu_\mathrm{ij}z'
	-2\pi T\sin\Delta\mu_\mathrm{ij}z' \right] e^{-2\pi T z'} \right\} \nonumber \\
 &=& F^a_\mathrm{ij}F^b_\mathrm{ji} \frac{N_c T^2}{2\pi}\left( e^{-\Delta\mu_\mathrm{ij}/T}-1\right) \left\{ -\frac{1}{z}\left[
	\frac{1}{\Delta\mu_\mathrm{ij}} \left( \bar{\mu}_\mathrm{ij} \sin \bar{\mu}_\mathrm{ij}z 
	+(2\pi T + z^{-1})\cos \bar{\mu}_\mathrm{ij}z \right) \right. \right. \nonumber \\
 &&{}+ \left. \sin \Delta\mu_\mathrm{ij}z +\frac{1}{z}\frac{1}{(\Delta\mu_\mathrm{ij})^2+(2\pi T)^2}( 
	\Delta\mu_\mathrm{ij}\cos \Delta\mu_{ij}z + 2\pi T \sin \Delta\mu_\mathrm{ij}z) \right]e^{-2\pi Tz}
 \label{eq:vapaayleinen} \\ 
 && {}+\Delta\mu_\mathrm{ij} \mathrm{Re} [ \Ei(-2\pi T z + i\bar{\mu}_\mathrm{ij}z) 
	+ \Ei(-2\pi T z + i\Delta\mu_\mathrm{ij}z) ]
	-2\pi T \mathrm{Im}\, \Ei(-2\pi Tz +i\Delta\mu_\mathrm{ij}z) \bigg\}. \nonumber
\end{eqnarray}
To better understand the long distance behavior of this correlator we can approximate the exponential integrals by
\begin{equation}
 \Ei(x) = \frac{e^x}{x}\left(1+\frac{1}{x} +\mathcal{O}(x^{-2})\right),
\end{equation}
which leads to our final expression
\begin{eqnarray}
 C_z[S^a, S^b] &=& F^a_\mathrm{ij}F^b_\mathrm{ji} \frac{N_c T^2}{2\pi}\left( 1 -e^{-\Delta\mu_\mathrm{ij}/T} \right)
	\frac{1}{z}e^{-2\pi Tz} \left[ \frac{1}{\Delta\mu_\mathrm{ij}} \left( \bar{\mu}_\mathrm{ij}\sin \bar{\mu}_\mathrm{ij}z 
	+(2\pi T + \frac{1}{z})\cos \bar{\mu}_\mathrm{ij}z \right) \right.  \nonumber \\
 &&{}+\frac{1}{z}\frac{2}{(\Delta\mu_\mathrm{ij})^2+(2\pi T)^2}( \Delta\mu_\mathrm{ij}\cos \Delta\mu_\mathrm{ij}z 
	+2\pi T \sin \Delta\mu_\mathrm{ij}z) \nonumber \\
 &&{}+\frac{\Delta\mu_\mathrm{ij}}{(2\pi T)^2+\bar{\mu}_\mathrm{ij}^2}\Big( 2\pi T \cos\bar{\mu}_\mathrm{ij}z
	-\bar{\mu}_\mathrm{ij}\sin\bar{\mu}_\mathrm{ij}z  \nonumber \\
 &&\left. \left. \hspace{1cm} {}+\frac{1}{z}\frac{1}{(2\pi T)^2+\bar{\mu}_\mathrm{ij}^2}\left[ 2\bar{\mu}_\mathrm{ij}2\pi T 
	\sin \bar{\mu}_\mathrm{ij}z -((2\pi T)^2-\bar{\mu}_\mathrm{ij}^2)\cos\bar{\mu}_\mathrm{ij}z \right] \right) \right].
\end{eqnarray}


\section{Computation of the potential}
\label{app:potential}

In this Appendix we show how to compute the static quark-antiquark potential in the effective theory. To make the notation lighter, we will assume below that we are computing with modes $\omega_n>0$. We need to introduce a point-splitting and study the modified correlator
\begin{equation}
 C(\mathbf{r},z) \equiv \int\!\dd^{d-1}\mathbf{R}\langle
	\phi_\mathrm{i}^\dagger(\mathbf{R},z)\chi_\mathrm{j}(\mathbf{R},z)
	\chi_\mathrm{j}^\dagger(\mathbf{r}/2,0)\phi_\mathrm{i}(-\mathbf{r}/2,0) \rangle.
\end{equation}
Going to the momentum space and performing the trivial integration over $\mathbf{R}$, this becomes
\begin{equation}
 C(\mathbf{r},z) = \int\! \frac{\dd^d p}{(2\pi)^d} \frac{\dd^d p'}{(2\pi)^d} \frac{\dd^d q'}{(2\pi)^d} \frac{\dd q_3}{2\pi}\langle 
	\phi_\mathrm{i}^\dagger(p)\chi_\mathrm{j}(\mathbf{p},q_3)
	\chi_\mathrm{j}^\dagger(q')\phi_\mathrm{i}(p') \rangle
	e^{i(q_3-p_3)z -i(\mathbf{p'}+\mathbf{q'})\cdot \mathbf{r}/2}.
\end{equation}

The leading order diagram of Fig.~\ref{fig:diagram_types}(a) consists of two free quark propagators,
\begin{eqnarray}
  C^{(0)}(\mathbf{r},z) &=& -2N_c \int\! \frac{\dd^d p}{(2\pi)^d} \frac{\dd q_3}{2\pi}
	\frac{-i}{M_\mathrm{i} -ip_3 +\frac{\mathbf{p}^2}{2p_{0\mathrm{i}}}}
	\frac{-i}{M_\mathrm{j} +iq_3 +\frac{\mathbf{p}^2}{2p_{0\mathrm{j}}}}
	e^{i(q_3-p_3)z -i\mathbf{p}\cdot \mathbf{r}} \nonumber \\
  &=& 2N_c \theta(z) \int\! \frac{\dd^2 p}{(2\pi)^2}\exp\left[ -\left( M_\mathrm{i} +M_\mathrm{j} 
	+\frac{\mathbf{p}^2}{2p_{0\mathrm{i}}} +\frac{\mathbf{p}^2}{2p_{0\mathrm{j}}} \right)z  
	-i\mathbf{p}\cdot \mathbf{r} \right] \nonumber \\
  &=& \theta(z)N_c \frac{\pnij}{\pi z} \exp \left[ -\Mij z -\frac{\pnij}{2z}\mathbf{r}^2 \right],
\end{eqnarray}
where $\pnij$ is given by Eq.~(\ref{eq:pnij_def}) and $\Mij \equiv M_\mathrm{i} +M_\mathrm{j}$. The high $T$ limit of this expression together with its complex conjugate correctly reproduces the approximate 4d result Eq.~(\ref{eq:vapaayleinen_appr}). In the isoscalar case we have checked that the exact free theory result Eq.~(\ref{eq:vapaaskalaari}) is recovered if we add the relativistic kinematic correction operator $-\nabla_\bot^2/8M^3$.

The NLO diagrams in Fig.~\ref{fig:diagram_types}(b,c) are computed the same way, the only complication being the more involved pole structure in $p_3,q_3$ integrations. For $A_0$ exchange the diagrams give
\begin{eqnarray}
 \ref{fig:diagram_types}(\textrm{b}) &=& \gE^2\CF\int\!\frac{\dd^d k}{(2\pi)^d}\frac{1}{k^2+\mE^2}\frac{1}{k_3^2} 
	\left( 1-e^{ik_3z}\right) C^{(0)}(\mathbf{r},z) \\
 \ref{fig:diagram_types}(\textrm{c}) &=& \gE^2\CF\int\!\frac{\dd^d k}{(2\pi)^d}
	\frac{e^{ i\mathbf{k}\cdot \mathbf{r}}}{k^2+\mE^2}\frac{1}{k_3^2} \left( 1-e^{ik_3z}\right)\left( 1-e^{-ik_3z}\right)
	C^{(0)}(\mathbf{r},z),
\end{eqnarray}
and similarly for $A_3$. The full NLO correction is then
\begin{eqnarray}
 C^{(1)}(\mathbf{r},z) &=& \gE^2\CF C^{(0)}(\mathbf{r},z) \int \!\frac{\dd^d k}{(2\pi)^d}
 \frac{1}{k^2+\mE^2}\frac{1}{k_3^2}\left(2-e^{ik_3z}-e^{-ik_3z}\right)\left(1+e^{ i\mathbf{k}\cdot \mathbf{r}}\right) \nonumber \\
&&{}- \frac{1}{k^2+\lambda^2}\frac{1}{k_3^2}\left(2-e^{ik_3z}-e^{-ik_3z}\right)\left(1-e^{ i\mathbf{k}\cdot \mathbf{r}} \right),
\end{eqnarray}
where $\lambda$ is again the infrared regulator in the gluon propagator.

The leading order term satisfies the equation
\begin{equation}
 \left[ \partial_z + \Mij -\frac{1}{2\pnij}\nabla^2_\mathbf{r}\right]C^{(0)}(\mathbf{r},z) =
	2N_c \delta(z)\delta^2(\mathbf{r}),
\end{equation}
whereas the similar equation for the first order term can be formulated as
\begin{equation}
 \left[ \partial_z + \Mij -\frac{\nabla^2_\mathbf{r}}{2\pnij}\right]C^{(1)}(\mathbf{r},z) =
	-\gE^2\CF \omega_n^{-2\epsilon} \mathcal{K}\left(\frac{1}{z\omega_n},\frac{\nabla_\mathbf{r}}{\omega_n},
	\frac{\mE}{\omega_n},\frac{\lambda}{\omega_n},\frac{\pnij}{\omega_n},\frac{\Mij}{\omega_n},
	\mathbf{r}\omega_n\right) C^{(0)}(\mathbf{r},z),
\end{equation}
where $\mathcal{K}$ is some dimensionless kernel. The right-hand side is already multiplied by $\gE^2$, so we need to keep only terms of order unity inside the kernel to get the potential to order $g^2$. According to the power counting rules of Sec.~\ref{subsec:matching}, we can then expand in the first two parameters, taking only the leading term into account. This gives
\begin{equation}
\left[ \partial_z + \Mij -\frac{\nabla^2_\mathbf{r}}{2\pnij}\right]C^{(1)}(\mathbf{r},z) \approx
	-\gE^2\CF \omega_n^{-2\epsilon} \mathcal{K}\left(0,0,\frac{\mE}{\omega_n},\frac{\lambda}{\omega_n},
	\frac{\pnij}{\omega_n},\frac{\Mij}{\omega_n},\mathbf{r}\omega_n\right) C^{(0)}(\mathbf{r},z),
\end{equation}
where
\begin{eqnarray}
 \lefteqn{ \mathcal{K}\left(0,0,\frac{\mE}{\omega_n},\frac{\lambda}{\omega_n},\frac{\pnij}{\omega_n},
	\frac{\Mij}{\omega_n},\mathbf{r}\omega_n\right) } \nonumber \\  
 &=& - \lim_{z\to\infty} \partial_z \int \!\frac{\dd^d k}{(2\pi)^d} \frac{1}{k_3^2}\left(2-e^{ik_3z}-e^{-ik_3z}\right) 
	\left[ \frac{1}{k^2+\mE^2} \left(1+e^{ i\mathbf{k}\cdot \mathbf{r}}\right)
	-\frac{1}{k^2+\lambda^2}\left(1-e^{ i\mathbf{k}\cdot \mathbf{r}} \right) \right] \nonumber \\
 &=& -\lim_{z\to\infty} \int \!\frac{\dd^d k}{(2\pi)^d} \frac{2\sin k_3 z}{k_3} 
	\left[ \frac{1}{k^2+\mE^2} \left(1+e^{ i\mathbf{k}\cdot \mathbf{r}}\right)
 	-\frac{1}{k^2+\lambda^2}\left(1-e^{ i\mathbf{k}\cdot \mathbf{r}} \right) \right] \nonumber  \\ 
 &=& -\int \!\frac{\dd^d k}{(2\pi)^d} 2\pi\delta(k_3) 	\left[ \frac{1}{k^2+\mE^2} \left(1+e^{ i\mathbf{k}\cdot \mathbf{r}}\right)
 	-\frac{1}{k^2+\lambda^2}\left(1-e^{ i\mathbf{k}\cdot \mathbf{r}} \right) \right] \nonumber  \\ 
 &=& \int \!\frac{\dd^{2-2\epsilon} \mathbf{k}}{(2\pi)^{2-2\epsilon}} 
	\left[ \frac{1}{\mathbf{k}^2+\lambda^2} \left(1-e^{ i\mathbf{k}\cdot \mathbf{r}}\right)
	-\frac{1}{\mathbf{k}^2+\mE^2}\left(1+e^{ i\mathbf{k}\cdot \mathbf{r}} \right) \right].
\end{eqnarray}
The signs here co-operate just right to give both ultraviolet ($\epsilon \to 0$) and infrared ($\lambda \to 0$) finite expression. The integrals needed here are
\begin{eqnarray}
\int \!\frac{\dd^{2-2\epsilon} k}{(2\pi)^{2-2\epsilon}}  \frac{1}{k^2+\lambda^2} &=&
	\frac{\mu^{-2\epsilon}}{4\pi}\left(\frac{1}{\epsilon} +2\ln\frac{\bar{\mu}}{\lambda} + \mathcal{O}(\epsilon)\right), \\
\int \!\frac{\dd^2 k}{(2\pi)^2}  \frac{e^{i\mathbf{k}\cdot \mathbf{r}}}{k^2+\lambda^2} &=&
	\frac{1}{2\pi}K_0(\lambda r) = \frac{1}{2\pi}\left( -\ln\frac{\lambda r}{2} -\gamma_E +\mathcal{O}(\lambda r)\right) ,
\end{eqnarray}
so we have
\begin{equation}
 V(\mathbf{r}) \equiv \gE^2\CF \mathcal{K}\left(0,0,\frac{\mE}{\omega_n},\frac{\lambda}{\omega_n},\frac{\pnij}{\omega_n},
 \frac{\Mij}{\omega_n},\mathbf{r}\omega_n\right) = \frac{\gE^2\CF}{2\pi}\left[ \ln \frac{\mE r}{2} +\gamma_E -K_0(\mE r)\right],
\end{equation}
and the full correlator $C(\mathbf{r},z)=C^{(0)}(\mathbf{r},z)+C^{(1)}(\mathbf{r},z)$ satisfies Eq.~(\ref{eq:corr_schrod}) as required. Note that the expansion in $1/z$ above corresponds to the static potential $t\to\infty$ in the corresponding 2+1 -dimensional theory. Similar methods have been recently used to compute the real-time static potential for a heavy quark-antiquark pair in 4d QCD \cite{Laine:2006ns}. The potential is infrared finite to this order, so we do not need to consider the contribution of the non-perturbative, long distance massless gauge fields.

\end{fmffile}

\bibliographystyle{h-physrev4}
\bibliography{/home/mtvepsal/texfiles/bibliography/articles.bib}

\end{document}